\documentclass[letterpaper,10pt,twocolumn,oneside,journal]{IEEEtran}

\usepackage{caption}

\usepackage{amsmath,amsfonts,amsthm,amssymb,graphicx,caption}
\usepackage{longtable}

\newcommand{\img}{\mathcal{I}}

\newcommand{\bzero}{\mathbf{0}}
\newcommand{\cov}{{\rm cov}}

\newcommand{\bone}{\boldsymbol{1}}

\newtheorem{definition}{Definition}[section]

\newcounter{theassumption} 
\newenvironment{assumption}{\par\noindent \underline{\textit{Assumption \Roman{theassumption}.}}\stepcounter{theassumption}\begin{itshape}}{\end{itshape}\par\noindent}
\newcounter{theassumptionsmall} 

\def\btheta{ { \theta} }

\def\vect{ {\rm vec} }

\def\cL{ {\mathcal L} }

\def\bmu{ { \mu} }

\begin{document}

\title{Cram\'{e}r-Rao Lower Bound for Point Based Image Registration with Heteroscedastic Error Model for Application in Single Molecule Microscopy}

\author{E.A.K.~Cohen, D.~Kim and
        R.J.~Ober
\thanks{This research was supported in part by the National Institute of Health grant R01 GM085575 and the Engineering and Physical Sciences Research Council (UK) mathematics platform grant EP/I019111/1.}
\thanks{E.A.K. Cohen is with the Department of Mathematics, Imperial College London, SW7 2AZ, UK (email: e.cohen@imperial.ac.uk) and R.J. Ober and D. Kim are with the Department of Biomedical Engineering, Texas A\&M University, College Station, Texas 77843-3120, USA (email: raimund.ober@tamu.edu).}
\thanks{First manuscript received October 02 2014. Resubmitted manuscript received April 17 2015. Revised manuscript received June 18 2015. Final manuscript received June 29 2015.}
\thanks{Copyright (c) 2010 IEEE. Personal use of this material is permitted. However, permission to use this material for any other purposes must be obtained from the IEEE by sending a request to pubs-permissions@ieee.org.}}

\maketitle

\begin{abstract}
The Cram\'{e}r-Rao lower bound for the estimation of the affine transformation parameters in a multivariate heteroscedastic errors-in-variables model is derived. The model is suitable for feature-based image registration in which both sets of control points are localized with errors whose covariance matrices vary from point to point. With focus given to the registration of fluorescence microscopy images, the Cram\'{e}r-Rao lower bound for the estimation of a feature's position (e.g. of a single molecule) in a registered image is also derived. In the particular case where all covariance matrices for the localization errors are scalar multiples of a common positive definite matrix (e.g. the identity matrix), as can be assumed in fluorescence microscopy, then simplified expressions for the Cram\'{e}r-Rao lower bound are given. Under certain simplifying assumptions these expressions are shown to match asymptotic distributions for a previously presented set of estimators. Theoretical results are verified with simulations and experimental data.
\end{abstract}


\IEEEpeerreviewmaketitle

\section{Introduction}
\label{intro}

\IEEEPARstart{I}{mage} registration is the process of overlaying two or more images of the same scene \cite{zitova03}. Image registration techniques can be divided into two categories; intensity-based registration where gray scale values are correlated between images, e.g. \cite{ashburner97} \cite{chumchob09}, \cite{myronenko10}, and feature-based registration, whereby correspondence between the two images is determined through the matching of distinct features common in both images e.g. \cite{liao10}, \cite{yasein09}. 

This project is motivated by an important problem in single molecule microscopy, a recent major advancement in fluorescence microscopy which allows individual fluorescently labeled molecules to be imaged using optical microscopy techniques and individually localized with accuracies in the very low nanometer range   \cite{shen11}, \cite{shen111}, \cite{wong11}. In a typical experiment two different proteins in a cell are labeled with different fluorescent markers. The biological information is obtained from the relationship between the two labeled sets of proteins. The imaging experiment consists of taking one exposure for each of the labeled proteins, often using two cameras, each equipped with a wavelength dependent optical filter to capture the emission of the fluorescence for the corresponding proteins. In this fashion we obtain two different images each displaying different aspects
of the sample. In order to analyze these images they need to be registered, as it cannot be assumed that the cameras are aligned to the degree that is necessary to guarantee the nanometer level accuracy which is required to obtain the appropriate information. Registration is typically achieved by incorporating fiducial markers, usually small nanometer size beads, into the sample whose fluorescent properties are such they can be imaged in both cameras. These fiducial markers can therefore serve as control points (CPs) for feature-based registration. The characterization of the registration errors is critical in assessing the deterioration of the localization accuracy of a single molecule due to the registration. A number of further single molecule microscopy experiments lead to the same underlying registration problem. One important such example arises
from the correction of drift in time lapse experiments.
\begin{figure}[t!]
  \centering
      \includegraphics[width=7cm]{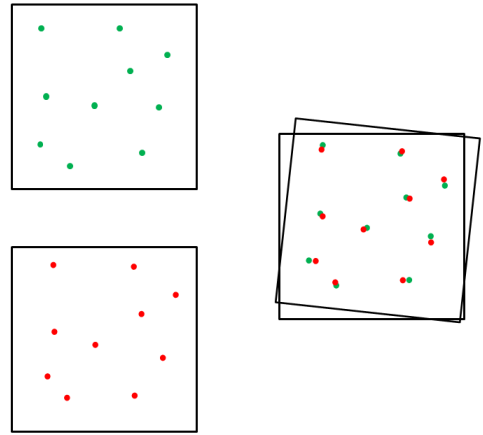}
  \caption{\label{realregplot}A diagram illustrating fiducial markers visible in both images (left). Each fiducial marker's position is located with a measurement error. Registration requires finding the transformation that best aligns the fiducial markers (right) with respect to an appropriate minimization problem.}
\end{figure}

Previous statistical studies on CP registration \cite{fitzpatrick98}, \cite{fitzpatrick98b} \cite{fitzpatrick01}, \cite{moghari06}, \cite{ma07}, \cite{wiles08}, \cite{moghari09}, \cite{ma10} assume one set of positions for the control points is taken as truth and errors  exist in only the second set of positions. In such examples a multivariate linear regression is used for the data model and the set of localization errors are sometimes referred to as the fiducial localization errors (FLEs) \cite{ma10}, \cite{moghari09}. In contrast, a recent study \cite{cohen13} presented a multivariate errors-in-variables (EIV) formulation of the control point image registration problem important to the microscopy application. The errors-in-variables formulation is necessary to model the situation when a ground truth for the CP locations is unavailable and the errors in measuring the CP locations are present in both images that are to be registered. 

Importantly, the microscopy application dictates that a heteroscedastic model is used as the measurement errors have to be assumed to have different covariance matrices for different CPs. A central aspect of the registration problem is the estimation of the registration transformation. There have been attempts in \cite{matei99}, \cite{matei00} to estimate registration parameters for heteroscedastic errors under an EIV model when the transformation is assumed rigid (rotation and translation only) with the heteroscedastic EIV (HEIV) algorithm; an iterative procedure that finds an optimal solution to the HEIV model. However, estimator distributions were only determined through bootstrapping methods. In the microscopy setting we take the more general assumption that the registration transformation be affine, allowed due to the high geometrical precision of modern microscope objectives. In \cite{cohen13} we have shown that for this data model a generalized maximum likelihood estimator is equivalent to a generalized least squares estimator. Using prior results we were able to obtain asymptotic results on the distributions for the estimators for the transformation parameters. For a specific heteroscedastic noise model where covariance matrices are scalar multiples of a known positive definite matrix, closed form expressions for estimators of the affine transformation parameters were derived. This particular model is applicable in a fluorescence microscopy setting where fiducial markers (e.g. fluorescent beads) act as the CPs but are each localized with differing degrees of accuracy. 

Registration performance is typically quantified by the fiducial registration error (FRE), which is the root mean-square distance between fiducial markers after registration, and most importantly the target registration error (TRE) which is the difference between corresponding points (other than the fiducial markers) after registration. The distribution of the TRE has been of much interest. Under the multivariate linear regression model (which as stated is inappropriate in fluorescence microscopy) \cite{fitzpatrick98}, \cite{fitzpatrick98b} derive approximate expressions for the root mean square of the TRE's absolute value and \cite{fitzpatrick01} gives its approximate distribution in the case where the registration transformation is assumed rigid (rotation and translation only) and FLEs are independent and identically distributed (iid) zero-mean Gaussian. Readers interested in the effect of biased FLEs are directed to \cite{moghari10}. Anisotropic iid FLEs are first considered in \cite{wiles08} and \cite{moghari09} derives the maximum likelihood estimators for the rigid transformation parameters along with the associated Cram\'{e}r-Rao Lower Bounds on their variance for this model. Heteroscedastic FLEs are considered in \cite{ma10} and using a spatial stiffness model they derive expressions for the root mean square TRE. An overview of these methods is given in \cite{shamir12}, together with procedures for the optimal selection of fiducial markers with respect to minimizing the TRE.

In \cite{cohen13} asymptotic distributions were found for the TRE under the multivariate errors-in-variables data model and affine transformation assumption required for fluorescence microscopy. Further to this, in \cite{cohen13} the asymptotic distribution was also found for the localization registration error (LRE), a newly defined measure of registration error that combines both a localization error and the TRE of a feature (e.g. single molecule) that is not used in the registration.

The quality of a single molecule experiment is assessed by the accuracy with which single molecules are localized in the particular experiment \cite{betzig06}.
Here the localization accuracy is interpreted as the standard deviation of an unbiased location estimator \cite{ober04}. In \cite{ober04} and \cite{ram12} the fundamental limit of localization accuracy was introduced as the Cram\'{e}r-Rao lower bound (CRLB) for the location estimation problem, in the context of ideal experimental conditions such as an infinite size photon detector without pixilation artefacts and without other extraneous noise sources. 
This measure has proved a reliable predictor for the best possible accuracy that can be achieved with a specific single molecule experiment \cite{abraham09}, \cite{lidke10}.

Due to the importance of registration in single molecule experiments the question therefore arises how the uncertainty introduced during the registration process influences the localization accuracy for a single molecule that has been registered. To this end, a major aspect of this manuscript consists of the derivation of the CRLB for the registration problem for several data models that are of relevance here.

The CRLB has been derived for registration problems before. The work of \cite{yetik06} and \cite{li09} consider the CRLB for feature-based and intensity-based registration performance in several scenarios of more general affine transformations between the two images, as well as a polynomial based non-linear transformation. However, they restrict themselves to the homoscedastic case, i.e. when all CP measurement errors have equal covariance matrix and consider only the CRLB of the transformation parameters themselves.

This paper provides the CRLB for registration performance when a general affine transformation is assumed and in the case of heteroscedastic CP measurement errors assumed zero-mean and Gaussian. We give particular focus to a fluorescence microscopy setting and not only consider the CRLB in estimating the transformation parameters, but place emphasis on finding the lower bound for the covariance matrix of the LRE, a concise and informative measure of registration performance. The square root of the diagonals of this covariance matrix (the standard deviation of the LRE in each dimension) is the accuracy with which single molecules are localized post-registration in each dimension. 

In Section \ref{formulation} we formulate the registration problem and define the LRE as introduced in \cite{cohen13}. In Section \ref{CRLBatp} we derive the CRLB for the affine transformation parameters in the most general heteroscedastic setting. In Section \ref{sml} we derive the CRLB for estimating the unknown position of a feature in the registered image, in turn giving a lower bound on the variance of the LRE. In Section \ref{wcgls} we consider the specific case when the covariance matrices of the measurement errors are a scalar multiple of a common matrix as is the case in a fluorescence microscopy setting. In fluorescence microscopy it is reasonable to approximate the covariance matrices as multiples of the identity matrix. Further, it is common for the affine transformation matrix to be a scalar multiple of a unitary matrix (a combination of scaling, rotation and reflection). In such a scenario we derive an explicit expression for the lower bound of the covariance matrix of the LRE that reveals a more intuitive view of these complex expressions. Importantly, this expression is identical to that of the asymptotic covariance of the LRE. This was derived in \cite{cohen13} under equivalent assumptions when the registration parameters are estimated using the corresponding generalized least-squares or equivalent maximum likelihood estimator. In Section \ref{simulations} we verify the theory with simulation studies and show the generalized least-squares estimator of \cite{cohen13} attains this lower bound. We conclude by considering real microscopy imaging data and show the CRLB results presented in this paper are appropriate in an experimental setting.

\section{Formulation}
\label{formulation}
We consider the registration experiment formulated in \cite{cohen13}. There are $K$ CPs located in both image 1, denoted $\img_{1}\subseteq\mathbb{R}^{d}$, and in image 2, denoted $\img_{2}\subseteq\mathbb{R}^{d}$ ($d=2$ or $3$). These CPs have true locations $\{x_{1,k}\in\img_1, k=1,...,K\}$ and $\{x_{2,k}\in\img_2, k=1,...,K\}$, respectively, and these CP coordinates are related by the affine transformation $T:\mathbb{R}^d\rightarrow\mathbb{R}^d$ where $x_{2,k} = T(x_{1,k}) = Ax_{1,k}+s$, $k=1,...,K$, with invertible $A\in\mathbb{R}^{d\times d}$ and $s\in\mathbb{R}^{d}$. The true positions of the CPs are not known in either image and instead must be measured with errors. We therefore observe the CP locations as $\{y_{1,k}\in\img_1, k=1,...,K\}$ and $\{y_{2,k}\in\img_2, k=1,...,K\}$, where $y_{j,k} = x_{j,k}+\epsilon_{j,k}$, $k=1,...,K$, $j=1,2$. The term $\epsilon_{j,k}\in \mathbb{R}^{d}$ is a random measurement error, sometimes referred to as the fiducial localization error (FLE), and are each assumed zero mean and to have individual covariance matrix $\Omega_{j,k}>0$ (where we use notation $M>0$ if matrix $M$ is positive definite and $M\geq 0$ if it is non-negative definite). All measurement errors are assumed to be pairwise independent across the CPs. 

Let us define the $\mathbb{R}^{d\times K}$ matrices $X_{j} \equiv \left[x_{j,1},...,x_{j,K}\right]$, $Y_{j} \equiv \left[y_{j,1},...,y_{j,K}\right]$ and $\mathcal{E}_{j}\equiv \left[\epsilon_{j,1},...,\epsilon_{j,K}\right]$, $j=1,2$. The measured control point locations can be conveniently represented as $Y_1 = X_1 + \mathcal{E}_1$ and $Y_2 = X_2 + \mathcal{E}_2$. The latter of these can be equivalently represented as $Y_2 = A X_1 + s\bone_{K}^{T} + \mathcal{E}_2$, where $^{T}$ is the matrix transpose and $\bone_{K}$ is a column vector of length $K$ with every element taking the value $1$.
If we further define the \emph{stacked} $\mathbb{R}^{2d\times K}$ matrices $X\equiv\left[X^{T}_{1},X^{T}_{2}\right]^{T}$, $Y\equiv\left[Y^{T}_{1},Y^{T}_{2}\right]^{T}$ and $\mathcal{E}\equiv\left[\mathcal{E}_{1}^{T},\mathcal{E}_{2}^{T}\right]^{T}$ then the system of equations can be condensed into the single matrix equation
\begin{equation}
\label{singleregression}
Y = \Lambda X_{1} + \alpha\bone_{K}^{T} + \mathcal{E},
\end{equation} 
where $\alpha = [\mathbf{0}^{T},s^{T}]^{T}$ and $\Lambda = [I_{d},A^{T}]^{T}$, with $I_d$ representing the $d$-dimensional identity matrix. The columns of $\mathcal{E}$ are independent with $k$th column $\epsilon_{k}\equiv[\epsilon_{1,k}^{T},\epsilon_{2,k}^{T}]^{T}$ having mean zero and known positive definite covariance matrix
\begin{equation}
\label{omegak}
\Omega_{k}\equiv\cov\{\epsilon_{k}\} = \left[\begin{array}{cc}\Omega_{1,k}&0\\0&\Omega_{2,k}\end{array}\right],
\end{equation}
where $\cov\{v\}$ denotes the covariance matrix of a random vector $v$. 

Models of type (\ref{singleregression}) are called errors-in-variables models. When covariance matrices $\{\Omega_{k},k=1,...,K\}$ all equal the same matrix $\Omega_0>0$ we have a homoscedastic errors-in-variables model. Under the homoscedastic assumption (\ref{singleregression}) is equivalent to the registration formulation of the CRLB study by \cite{yetik06} and \cite{li09}. When the $K$ covariance matrices $\{\Omega_{k},k=1,...,K\}$ are in general not equal then we have a heteroscedastic errors-in-variables model. It is the heteroscedastic assumption that this study focuses on.

Image registration requires estimating the transformation parameters $A$ and $s$ whose elements we can represent in the transformation parameter vector $\theta_T = [\vect(A)^T,s^T]^T$. In the strict homoscedastic case \cite{gleser81} defines the generalized least squares (GLS) estimators of $A$ and $s$ and shows them to be equivalent to the maximum likelihood (ML) estimators under the assumption of CP measurement errors being Gaussian. Further to this, closed form expressions for the estimators of $A$ and $s$ are given along with their joint and marginal asymptotic distributions. The work of \cite{chan84} considers the most general heteroscedastic model, where under the assumption of Gaussian measurement errors the maximum likelihood estimators for $A$ and $s$ are presented along with an iterative method for their computation and their joint and marginal asymptotic distributions. Recently in \cite{cohen13}, a heteroscedastic generalized least squares estimator is defined in an extension to the homoscedastic formulation of \cite{gleser81} and is shown to be equivalent to the maximum likelihood estimator considered in \cite{chan84}. In the special case where covariance matrices for the measurement errors are of the form $\Omega_k =\eta_k \Omega_0$, where $\eta_k\in\mathbb{R}^+$ and $\Omega_0>0$ --- termed the \emph{weighted covariance} model --- then \cite{cohen13} gives closed form expressions for the estimators of $A$ and $s$ and determines their joint and marginal asymptotic distributions. This in turn is used to give concise expressions for the first and second moment of the TRE and LRE, measures of registration error that we now formally define.
\subsection{Registration Errors}
\label{LRE}
\begin{figure}[t!]
  \centering
      \includegraphics[width=7cm]{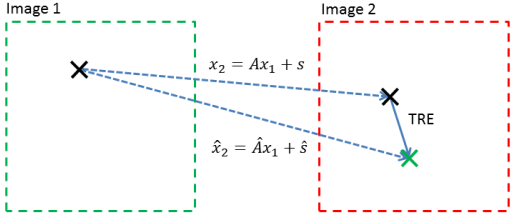}
  \caption{\label{TREplot}A diagram illustrating the target registration error as the difference between the true position of an arbitrary point in $\img_2$ (black/upper cross) and its registered position using the estimated registration parameters (green/lower cross).}
\end{figure}

\begin{figure}[t!]
  \centering
      \includegraphics[width=7cm]{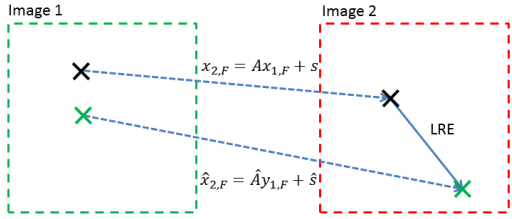}
  \caption{\label{LREplot}A diagram illustrating the localization registration error as the difference between the true position of feature in $\img_2$ (black/upper cross) and its registered position using its estimated position in $\img_1$ (green/lower cross) and the estimated registration parameters.}
\end{figure}

As has been stated in Section \ref{intro}, the TRE is a commonly used measure of registration performance. Here we give its definition when the registration transformation $T$ is assumed affine with matrix parameter $A$ and vector parameter $s$ (see Figure \ref{TREplot}). 
\begin{definition}
\label{tredef}
Let $A$ and $s$ be the registration transformation parameters and let $\hat{A}$ and $\hat{s}$ be their respective estimators. The target registration error (TRE) $\tau:\img_{1}\rightarrow\mathbb{R}^{d}$ for an arbitrary point $x_{1}\in\img_{1}$ with corresponding mapped position in $\img_2$ of $x_{2} = Ax_{1}+s$ is defined as
$
\tau(x_{1}) \equiv x_{2} - (\hat{A}x_{1} + \hat{s}) =  Ax_{1}+s - (\hat{A}x_{1} + \hat{s}).
$
\end{definition}
The LRE is defined in \cite{cohen13} and is of particular use in fluorescence microscopy registration experiments. Suppose we have a feature (e.g. a single molecule) that is visible in image 1 but not in image 2 (and therefore is not involved in the registration process). The LRE gives the error with which it is localized in image 2 after registration (see Figure \ref{LREplot}).
\begin{definition}Let $A$ and $s$ be the registration transformation parameters and let $\hat{A}$ and $\hat{s}$ be their respective estimators. For a feature (e.g. single molecule) in $\img_{1}$ with true and measured locations $x_{1,F}$ and $y_{1,F}=x_{1,F}+\epsilon_{1,F}$ respectively, the localization registration error (LRE) $\ell_{F}$ is defined as the difference between the true position in $\mathcal{I}_2$, given by $x_{2,F} = Ax_{1,F}+s$, and the registered position $\hat{x}_{2,F} = \hat{A}y_{1,F} + \hat{s}$, i.e.
$
\ell_{F} \equiv x_{2,F} - \hat{x}_{2,F}.
$
\end{definition}
The standard deviation of an element of $\ell_F$ is the accuracy with which a feature/single molecule can be localized in that dimension post-registration. The covariance matrix for $\ell_{F}$, denoted $\Omega_{\ell}$, is identical to the covariance matrix for the estimator $\hat{x}_{2,F}$ and therefore the CRLB for estimating $x_{2,F}$ is a lower bound for $\Omega_{\ell}$, i.e. if we denote the CRLB matrix for estimating $x_{2,F}$ as $C_{FF}$ then $\Omega_{\ell}\geq C_{FF}$ (where notation $\Omega_{\ell}\geq C_{FF}$ means $\Omega_{\ell}-C_{FF}\geq 0$, i.e. $\Omega_{\ell}-C_{FF}$ is non-negative definite). Further discussion on the relationship between the TRE and LRE can be found in \cite{cohen13}.

\section{CRLB for affine transformation parameters}
\label{CRLBatp}
The vector of unknown parameters is given as $\btheta_{TC} \equiv [\btheta_T^T,\btheta_C^T]^T$, where $\btheta_T \equiv  [\vect(A)^T,s^T]^T$ is the $\mathbb{R}^{(d^2 +d)}$ vector of affine transformation parameters and $\btheta_C \equiv [x_{1,1}^T,x_{1,2}^T,...,x_{1,K}^T]^T$ is the $\mathbb{R}^{dK}$ vector of CP location parameters. We make the assumption that measurement errors are independent, are zero-mean (validated in \cite{abraham09}) and are Gaussian (validated in Section \ref{simulations}) with covariance matrix $\Omega_k$ of form (\ref{omegak}), i.e. $\epsilon_{k}\sim N_{2d}(\bzero,\Omega_k)$. The likelihood function is therefore given as \cite{cohen13},\cite{chan84}
\begin{multline*}
L(\btheta_{TC}|y_{1},...,y_{K}) = \frac{1}{(2\pi)^{Kd/2}}\left(\prod_{k=1}^{K}\left|\Omega_{k}\right|^{-1/2}\right)\\ \times \exp\left(-\frac{1}{2}\sum_{k=1}^{K}(y_k-\mu_k)^T\Omega_{k}^{-1}(y_k-\mu_k)\right),
\end{multline*}
where $\mu_k = [x_{1,k}^T,x_{2,k}^T]^T$, $|\Omega_k|$ denotes the determinant of $\Omega_k$ and $y_1,...,y_K$ are the stacked vectors of measured CP locations. The corresponding log-likelihood is
\begin{multline*}
\cL(\btheta_{TC}|y_{1},...,y_{K}) = \\ -\frac{Kd}{2}\ln (2\pi) - \frac{1}{2} \sum_{k=1}^{K}\ln|\Omega_{k}|-\frac{1}{2}\sum_{k=1}^{K}(y_k-\bmu_k)^T\Omega_{k}^{-1}(y_k-\bmu_k).
\end{multline*}
It is well established that for the multivariate normal distribution with covariance that is independent of parameters the Fisher information matrix (FIM) for the parameter vector $\theta_{TC}$, denoted $J(\theta_{TC})$, is given as \cite[p. 47]{kay93}
\begin{equation}
\label{JTC}
J(\theta_{TC}) = \sum_{k=1}^{K}\frac{\partial \bmu_k^T}{\partial \btheta_{TC}}\Omega_k^{-1}\frac{\partial \bmu_k}{\partial \btheta_{TC}^T},
\end{equation}
and the CRLB matrix is given as $C(\theta_{TC}) \equiv J^{-1}(\theta_{TC})$. With the CRLB matrix denoted as
\begin{equation}
C(\btheta_{TC}) = \left[\begin{array}{cc}C_{TT} & C_{TC} \\ C_{CT} & C_{CC}\end{array}\right],\label{cthetatc}
\end{equation}
the diagonals of $C_{TT}$ are the CRLBs for estimating the \emph{transformation} parameters and the diagonals of $C_{CC}$ are the CRLBs for estimating the \emph{control point} locations. We are therefore primarily interested in the diagonals of $C_{TT}$. It is shown in Appendix \ref{CRLBTT} that
$$
J(\theta_{TC}) =\sum_{k=1}^{K} \left[\begin{array}{cc}H_k^T\Omega^{-1}_{2,k}H_k & H^T_k\Omega_{2,k}^{-1}G_k \\ G^T_k\Omega_{2,k}^{-1}H_k & F_k^T\Omega_{1,k}^{-1}F_k + G_k^T \Omega_{2,k}^{-1}G_k \end{array}\right],
$$
where $F_k  = (e_K^{(k)})^T\otimes I_d$, $G_k = (e_K^{(k)})^T\otimes A$, $H_k = \left[I_d\otimes x_{1,k}^T,I_d\right]$. Here, $e^{(j)}_p$ represents the $j$th standard basis vector of $\mathbb{R}^{p}$, (i.e. vector of length $p$ with 1 placed in the $j$th entry and zeros everywhere else) and $\otimes$ denotes the Kronecker product. It follows from the block matrix inversion of $J(\theta_{TC})$ that
\begin{equation}C_{TT} = \left(S_{HH} - S_{HG}\left(S_{FF}+S_{GG}\right)^{-1}S_{HG}^T\right)^{-1},\label{eqCRLBTT}\end{equation}
where $S_{HH} = \sum_{k=1}^{K}H_k^T\Omega_{2,k}^{-1}H_k$, $S_{HG} = \sum_{k=1}^{K}H^T_{k}\Omega_{2,k}^{-1} G_k$, $S_{FF} = \sum_{k=1}^{K}F_k^T\Omega^{-1}_{1,k}F_k$ and $S_{GG} = \sum_{k=1}^K G_k^T\Omega_{2,k}^{-1}G_k$. This is the CRLB for the transformation parameters in the most general heteroscedastic errors-in-variables model considered in \cite{chan84}. Equations (\ref{HH}), (\ref{HG}) and (\ref{FFGG}) in Appendix \ref{CRLBTT} provide expressions for the sums $S_{HH}$, $S_{HG}$ and $S_{FF} + S_{GG}$, respectively.

\section{Feature localization}
\label{sml}
Let us now consider including the localization of a feature (e.g. single molecule) into the expression. For this we include the term $y_{1,F}$ - the observed location of the feature in $\mathcal{I}_1$, and the unknown parameter $x_{2,F}$ - the true position of the feature in $\mathcal{I}_2$ that we wish to estimate. The associated localization error has covariance matrix $\Omega_{1,F}$. As previously stated in Section \ref{LRE}, the LRE $\ell_F\in\mathbb{R}^d$ is the difference between the estimator $\hat{x}_{2,F}=\hat{A}y_{1,F} + \hat{s}$ and the true value $x_{2,F}$ and hence the CRLB for estimating $x_{2,F}$ provides a lower bound for $\Omega_{\ell}$, the covariance matrix of $\ell_F$.

The combined log-likelihood function for all parameters $\theta_{FTC} \equiv [\theta_F^T,\theta_T^T,\theta_C^T]^T$, where $\theta_F \equiv x_{2,F}$, given the observed data is now
\begin{multline}
\mathcal{L}(\btheta_{FTC}|y_{1},...,y_{K},y_{1,F}) =-\frac{d}{2}\ln(2\pi)-\frac{1}{2}\ln \left|\Omega_{1,F}\right|\\-\frac{1}{2}(y_{1,F}-\mu_{1,F})^T\Omega_{1,F}^{-1}(y_{1,F}-\mu_{1,F}) -\frac{Kd}{2}\ln (2\pi)\\ - \frac{1}{2} \sum_{k=1}^{K}\ln|\Omega_{k}|-\frac{1}{2}\sum_{k=1}^{K}(y_k-\bmu_k)^T\Omega_{k}^{-1}(y_k-\bmu_k).\label{loglike}
\end{multline}
In terms of the unknown parameter $\theta_F \equiv x_{2,F}$ we can write $\mu_{1,F} = A^{-1}(x_{2,F}-s)$. The FIM for the complete parameter vector $\theta_{FTC}$ is shown in Appendix \ref{FIMLRE} to be given as
\begin{multline}
\label{eqFIMLRE}
J(\theta_{FTC}) = \\ \left[\begin{array}{ccc}A^{-T}\Omega_{1,F}^{-1}A^{-1} & D_{FT} & 0  \\ D_{FT}^T & D_{TT}+ S_{HH} & S_{HG} \\ 0 & S_{HG}^T & S_{FF} + S_{GG} \end{array}\right],
\end{multline}
where $D_{TT} \equiv D_T^T\Omega_{1,F}^{-1}D_T$ and $D_{FT} \equiv D_F^T\Omega_{1,F}^{-1}D_T$, with
\begin{align}
D_F\equiv \frac{\partial \bmu_{1,F}}{\partial \btheta^T_F} & =  -A^{-1},\nonumber\\
D_T \equiv \frac{\partial \bmu_{1,F}}{\partial \btheta^T_T} & = -A^{-1}\left[x_{1,F}^T\otimes I_d,I_d\right].\nonumber
\end{align}
Representing the inverse FIM of $\theta_{FTC}$ as
$$
C(\theta_{FTC})\equiv J^{-1}(\theta_{FTC}) = \left[\begin{array}{ccc}
C_{FF} & C_{FT} & C_{FC} \\
C_{TF} & C_{TT} & C_{TC} \\
C_{CF} & C_{CT} & C_{CC}
\end{array}\right],
$$
it is shown in Appendix \ref{CRLBLRE} that the sub-block
$
\left[\begin{smallmatrix}
C_{TT} & C_{TC} \\
C_{CT} & C_{CC}
\end{smallmatrix}\right]
$
is identical to $C(\theta_{TC})$ in (\ref{cthetatc}) (as one would expect from the fact that the feature/single molecule is not involved in the registration process), and the CRLB matrix for estimating $x_{2,F}$, the location of a feature/single molecule in the registered image, is given by
\begin{equation}
\label{eqCRLBLRE}
C_{FF} = \left(A^{-T}\Omega_{1,F}^{-1}A^{-1} - D_{FT}(D_{TT} + C_{TT}^{-1})^{-1}D_{FT}^T\right)^{-1},
\end{equation}
where $C_{TT}$ is the CRLB matrix for estimating the transformation parameters given in (\ref{eqCRLBTT}). The $d$ diagonal elements of $C_{FF}$ are the CRLBs for estimating the respective elements of $x_{2,F}$, and with $\Omega_{\ell}\geq C_{FF}$ offers the lower bounds on the variances of the respective elements of the LRE $\ell_F$.  
\section{CRLB expressions for weighted covariance model}
\label{wcgls}
Section \ref{sml} provides the CRLB for localizing a feature/single molecule in the most general heteroscedastic registration model. While these results provide a very general solution to our problem, we will now investigate special cases that are of interest in their own right through their relevance in applications. In addition, in these special cases we can obtain significant simplifications of the above expressions that provide useful insights for experimental design considerations. In this section we look to the \emph{weighted covariance} model formulated in \cite{cohen13}, in which we make the assumption that covariance matrices for the measurement errors are of the form $\Omega_k =\eta_k \Omega_0$ where $\eta_k\in\mathbb{R}^+$ and $\Omega_0>0$, for all $k=1,...,K$. Here, we consider the following further assumption.
\begin{assumption}
\label{A2}
Covariance matrices have the forms: $\Omega_{1,0} = \sigma_{1,0}^2 I_2$, $\Omega_{2,0} = \sigma_{2,0}^2 I_2$, $\Omega_{1,F} = \sigma_{1,F}^2 I_2$, and transformation matrix $A = \varsigma R$, where $R$ is a unitary matrix (rotation/reflection) and $\varsigma \in \mathbb{R}^+$ is a scaling factor.
\end{assumption} The transformation vector $s$ is arbitrary. The assumption here that the covariance matrices are some scalar multiple of the identity matrix is a reasonable assumption in fluorescence microscopy and exact in the case of a non-pixelated detector \cite{ober04}. The assumption on the transformation matrix is a common type of transform experienced in registration.
\subsection{CRLB for estimating transformation parameters}
Let us define the following quantities that will be used here: $\gamma \equiv (1/K)\sum_{k=1}^K\eta_k^{-1}$, $\chi_k \equiv x_{1,k}x_{1,k}^T$ ($k=1,...,K$), $\Xi \equiv (1/K)\sum_{k=1}^K \eta_{k}^{-1}\chi_k$, $X_{i,k} \equiv e_{2}^{(i)}\otimes x_{1,k}^T$ ($i=1,2$ and $k=1,...,K$), $\bar{x}_1 \equiv (1/K)\sum_{k=1}^{K}\eta_{k}^{-1}x_{1,k}$, $\bar{X}_{i} \equiv e_{2}^{(i)}\otimes \bar{x}_{1}^T$ ($i=1,2$), $\Psi \equiv \Xi - \gamma^{-1}\bar{x}_1\bar{x}_1^T$ and $\Gamma_i \equiv \gamma^{-1}\bar{X}_i\Psi$ ($i=1,2$).

Under Assumption I it is shown in Appendix \ref{simplify} that the expression for $C_{TT}$ in (\ref{eqCRLBTT}) simplifies to
\begin{multline}C_{TT} = \frac{1}{K}\left(\varsigma^2\sigma_{1,0}^{2} + \sigma_{2,0}^{2}\right)\\ \times\left[\begin{array}{ccc}
\Psi^{-1} & 0 & -\Gamma_1^T \\
0 & \Psi^{-1} & -\Gamma_2^T \\
-\Gamma_1 &  -\Gamma_2 & \gamma^{-1} I_2 + \gamma^{-1}\left(\Gamma_1\bar{X}_1^{T}+\Gamma_2\bar{X}_2^{T}\right)
\end{array}\right].\label{CTTeq}
\end{multline}

\subsection{CRLB for estimating the location of a feature/single molecule in the registered image}
\subsubsection{General model}
Under Assumption I it is shown in Appendix \ref{newapp} that the CRLB matrix for estimating the location $x_{2,F}$ of a feature/single molecule is given as
\begin{multline}
C_{FF} = \left(\frac{1}{\varsigma^2\sigma_{1,F}^2}I_2 - \right. \\ \left. \frac{1}{\varsigma^4\sigma_{1,F}^4}\left[x_{1F}^T,1\right]\otimes I_2\left(\frac{1}{\varsigma^2\sigma_{1,F}^2}\left[\begin{array}{cc}
x_{1,F}x_{1,F}^T & x_{1,F} \\ 
x_{1,F}^T & 1
\end{array} \right]\otimes I_2 \right.\right.\\ \left.\left. + \left(\varsigma^2\sigma_{1,0}^{2} + \sigma_{2,0}^{2}\right)^{-1}\sum_{k=1}^{K}\eta_{k}^{-1}\left[\begin{array}{ccc}
\chi_k & 0 & X_{1,k}^T \\ 0 & \chi_k & X_{2,k}^T \\ X_{1,k} & X_{2,k} & I_2
\end{array}\right]\right) ^{-1} \right. \\ \left. \times \left[x_{1F}^T,1\right]^T\otimes I_2\right)^{-1}.\label{CFFeq}
\end{multline}

\subsubsection{Simplified model}
Let us consider the case where $\eta_k$ is independent of CP position and CPs are centrally and symmetrically distributed in the image space. This model leads to the following set of assumptions that are appropriate for large $K$ and asymptotically exact. These assumptions naturally arise, for example, when considering the experimental disposition of fluorescent beads in a specific microscopy experiment setting, where the beads can be assumed to have a circular Gaussian spatial distribution \cite{cohen13}.
\begin{assumption}
Approximate $(1/K)\sum_{k=1}^{K}\eta_k^{-1}\chi_k = \nu^2 I_2$ and $(1/K)\sum_{k=1}^{K}\eta_k^{-1}X_{j,k} = 0$.
\end{assumption}
Under Assumption I and II it is shown in Appendix \ref{CRLBsupersimple} that the CRLB matrix for estimating the feature/single molecule location $x_{2,F}$ is given as
$$
C_{FF} = \varsigma^2\sigma^2_{1,F}I_2 + \frac{1}{K\gamma}\left(\varsigma^2\sigma_{1,0}^{2} + \sigma_{2,0}^{2}\right)\left(1+\frac{\gamma r^2}{\nu^2}\right)I_2.
$$
\subsubsection{Applying to a microscopy setting}
\label{atams}
Consider a fluorescence microscopy example were Assumptions I and II are satisfied. That is, we have the weighted covariance model with $\eta_k = N_{1,k}^{-1}$, where $N_{1,k}$ is the number of photons associated with control point $k$ in $\mathcal{I}_1$, $k = 1,...,K$, and $\Omega_{j,0} =\sigma_{j,0}^2I_2$ where $\sigma_{1,0}^2 = \zeta_1$ and $\sigma_{2,0}^2 = \zeta_2/c$ \cite{cohenas12}. Constant $\zeta_j$, $j=1,2$, is a known localization accuracy parameter associated with $\mathcal{I}_j$ and is a function of the numerical aperture, photon wavelength and point spread function (see \cite[p. 6296]{cohen13}). Constant $c$ is the constant of proportionality assumed in \cite{cohen13} to exist such that $N_{2,k} = cN_{1,k}$ where $N_{2,k}$ are the number of photons associated with the $k$th CP in $\mathcal{I}_2$. This gives
$$
\Omega_k = \frac{1}{N_{1,k}}\left(\begin{array}{cc}\zeta_1 I_2 & 0 \\ 0 & c^{-1}\zeta_2 I_2\end{array}\right).
$$ 
In this situation we have $\gamma = \bar{N}_1$ where $\bar{N}_1$ is the mean photon count for the CPs in $\mathcal{I}_1$.
Therefore the CRLB matrix for estimating $x_{2,F}$, the location of the single molecule in $\mathcal{I}_{2}$, is given as
$$
C_{FF} = \varsigma^2\sigma^2_{1,F}I_2 + \frac{1}{K}\left(\varsigma^2\frac{\zeta_{1}}{\bar{N}_1} + \frac{\zeta_{2}}{\bar{N}_2}\right)\left(1+\frac{\bar{N}_1 r^2}{\nu^2}\right)I_2.
$$
Photon counts are independent of CP position and therefore under the (asymptotically exact \cite{cohen13}) assumption that $\nu^2 = \bar{N}_1\kappa^2$, where $(1/K)\sum_{k=1}^K \chi_k = \kappa^2I_2$ (a measure of the spread of the CPs), then
\begin{equation}
\label{finaleq}\Omega_{\ell}\geq C_{FF} = \varsigma^2\sigma^2_{1,F}I_2 + \frac{1}{K}\left(\varsigma^2\frac{\zeta_{1}}{\bar{N}_1} + \frac{\zeta_{2}}{\bar{N}_2}\right)\left(1+\frac{r^2}{\kappa^2}\right)I_2.
\end{equation}
This expression for the CRLB in estimating $x_{2,F}$, and hence the lower bound for $\Omega_{\ell}$, exactly matches the large $K$ expression for $\Omega_{\ell}$ found in \cite[p. 6297]{cohen13} when the generalized least squares estimator for the weighted covariance model is used. The fundamental limit (i.e. the theoretical lower bound) of localization accuracy for a single molecule in a pair of registered images is therefore bound by a term that depends on $K$ (the number of CPs used in the registration process) and their associated photon counts, along with $\kappa^2$ that gives a measure of the spread of the CPs in the image. We note there is no dependence on the translation parameter $s$.

\section{Simulations and experimental verification}
\label{simulations}
In this section we verify the CRLB results given in this paper with computational simulations and a real data experiment. 
\subsection{Simulation Studies}
\label{s1}
In these computational simulation studies we consider a microscopy experiment where we register a pair of different coloured monochromatic images each captured using an optical system with identical numerical aperture and point-spread function. The measurement error in localizing the $k$th CP in $j$th image $\mathcal{I}_j$ ($j=1,2$) has zero mean and covariance matrix $(\zeta_{j}/N_{j,k})I_2$ where $\zeta_j=\lambda^2_{j,em}/(4\pi^2 n_F^2)$ \cite{ober04}. The photon wavelength $\lambda_{j,em}$ associated with each image is 540nm and 650nm respectively, $n_F$ is the numerical aperture and assigned a typical value of 1.4 and $N_{j,k}$ is the photon count associated with the $k$th control point in the $j$th image.
\subsubsection{Rotation}
\label{rotation}
CPs are arranged in a square grid of side length 81$\mu$m in the object space with varying numbers of points within that grid, and therefore $K$ is restricted to the square numbers from 4 to 64. The photon counts associated with each control point are observed realisations of a uniformly distributed random integer on the interval [5000,10000]. In $\mathcal{I}_1$ is a single molecule at position (16$\mu$m,20$\mu$m) from the center, with which a photon count of 1000 is associated. Affine transformation matrix $A$ is a rotation matrix of angle $30$ degrees and affine transformation vector $s$ is [4.8$\mu$m,4.8$\mu$m]$^T$. 

\begin{figure}[h]
\begin{center}
\includegraphics[width=80mm]{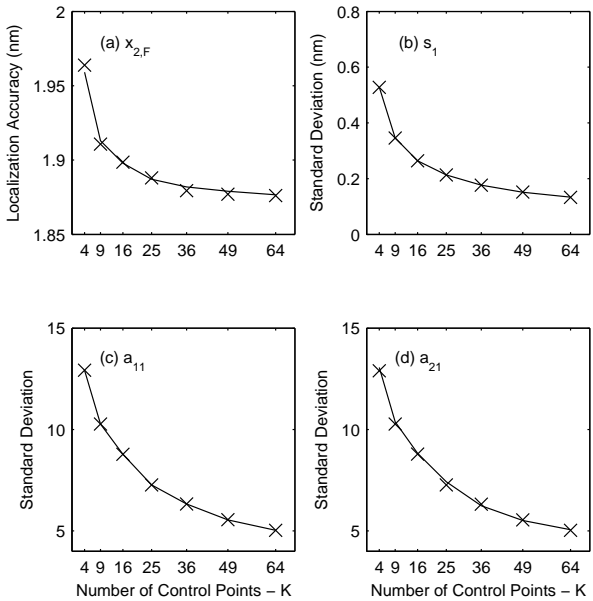}
\end{center}
 \caption{\label{figsRotation}The line indicates the square root of the CRLB and the crosses mark the sample standard deviation plotted as a function of the number of CPs $K$ for the following estimated parameters: (a) the first dimension of the unknown single molecule position parameter $x_{2,F}$, (b) $s_1$, the first element of transformation vector $s$, (c) $a_{11}$, the leading element of transformation matrix $A$, (d) $a_{21}$, the $(2,1)$th element of $A$. In this simulation study CPs are in a grid configuration and the transformation consists of a rotation and translation (see Section \ref{rotation} for more details). The vertical axes in (a) and (b) have units as nanometers, the vertical axes in (c) and (d) are unitless and are on the scale $\times 10^{-5}$. Results are based on a sample of $10^6$ simulations.}
\end{figure}

We look to verify the CRLB for estimating the transformation parameters as given in (\ref{CTTeq}) and the CRLB for estimating the position of the single molecule in the registered image (equivalently the lower bound of $\Omega_{\ell}$) in (\ref{finaleq}). This is achieved by estimating the transformation parameters using the generalized least squares estimator for the weighted covariance model as developed in \cite{cohen13}. The empirical standard deviations of interest are computed using $10^6$ simulations and shown in Figure \ref{figsRotation}.
\subsubsection{Shear}
\label{shear}
CPs are arranged in a square grid of side length 81$\mu$m in the object space with $K=9$. The photon counts associated with each control point are observed realisations of a uniformly distributed random integer on the interval [5000,10000] and covariance matrices for the measurement errors are of the same form as in Section VI-A1. In $\mathcal{I}_1$ is a single molecule at position (16$\mu$m,20$\mu$m) from the center, with which a photon count of 1000 is associated. Affine transformation matrix $A$ is a shear matrix of type
$
A = \left(\begin{smallmatrix}
1 & \lambda \\ 0 & 1
\end{smallmatrix}\right)
$
where shear parameter $\lambda$ is varied between values of 0.1 and 0.9. Transformation vector $s$ is [4.8$\mu$m,4.8$\mu$m]$^T$. 

\begin{figure}[h]
\begin{center}
\includegraphics[width=80mm]{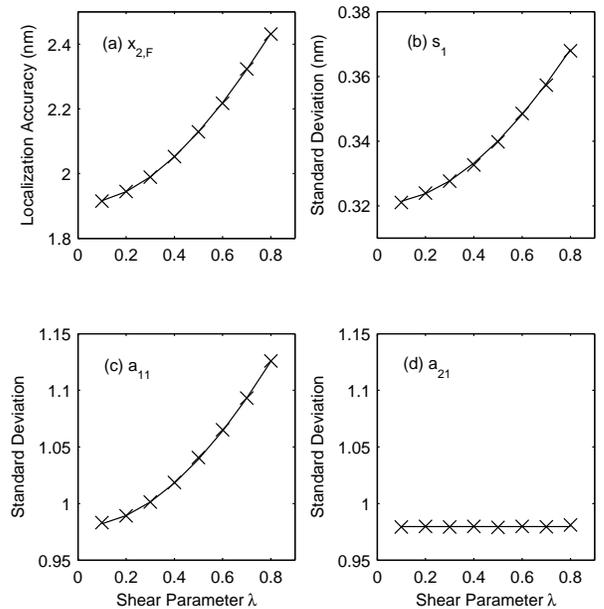}
 \caption{The line indicates the square root of the CRLB and the crosses mark the sample standard deviation plotted as a function of shear parameter $\lambda$ for the following estimated parameters: (a) the first dimension of the unknown single molecule position parameter $x_{2,F}$, (b) $s_1$, the first element of transformation vector $s$, (c) $a_{11}$, the leading element of transformation matrix $A$, (d) $a_{21}$, the $(2,1)$th element of $A$. In this simulation study there are nine CPs in a grid configuration (see Section \ref{shear} for more details). The vertical axes in (a) and (b) have units as nanometers, the vertical axes in (c) and (d) are unitless and are on the scale $\times 10^{-4}$. Results are based on a sample of $10^6$ simulations.}
\label{figsShear}
\end{center}
\end{figure}

We look to verify the CRLB for estimating the transformation parameters as given in (\ref{CTTeq}) and the CRLB for estimating the position of the single molecule in the registered image (equivalently the lower bound of $\Omega_{\ell}$) in (\ref{finaleq}). This is achieved by estimating the transformation parameters using the generalized least squares estimator for the weighted covariance model as developed in \cite{cohen13}. The empirical standard deviations of interest are computed using $10^6$ simulations and shown in Figure \ref{figsShear}.
\subsubsection{Asymptotic covariance versus CRLB}
It has been mentioned in Section \ref{atams} that under Assumption I and II the lower bound for $\Omega_{\ell}$ in (\ref{finaleq}) matches the large $K$ covariance matrix of the LRE given in \cite{cohen13} when transformation parameters are estimated using the generalized least squares estimator. We now consider relaxing Assumption I such that $\Omega_0$ is no longer the identity matrix and look at how the CRLB for estimating $x_{2,F}$ compares with the more general large $K$ covariance matrix expression in \cite[p. 6295]{cohen13}.  

\begin{figure}[t!]
  \centering
      \includegraphics[width=5cm]{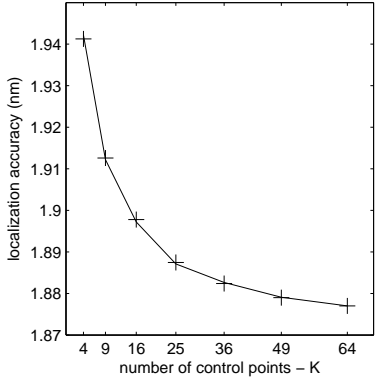}
  \caption{\label{noniid}The line indicates the square root of the CRLB for the first dimension of the unknown single molecule position parameter $x_{2,F}$ (and hence a lower bound for the standard deviation of the LRE) in object space dimensions, plotted as a function of the number of CPs. CPs are in a grid configuration (see Section \ref{simulations} for more details). The crosses are the theoretical standard deviation of the LRE assuming the large $K$ distribution given in \cite{cohen13}.}
\end{figure}

We have exactly the same experimental set-up as in Section \ref{rotation} except the measurement error in localizing the $k$th CP in $j$th image $\mathcal{I}_j$ ($j=1,2$) now has covariance matrix $(\zeta_{j}/N_{j,k})S$ where $S = \left(\begin{smallmatrix}
1 & 0.5 \\ 0.5 & 1
\end{smallmatrix}\right).$ The CRLB for estimating the single molecule location $x_{2,F}$ is calculated using the more general expression (\ref{eqCRLBLRE}). In Figure \ref{noniid} the square root of its leading diagonal is compared to the large $K$ standard deviation for the first dimension of the LRE given in \cite[p. 6295]{cohen13} when registration is performed using the generalized least squares estimator. It is clear to see that the two expressions take very similar values, particularly for large values of $K$, and hence the close association between the CRLB expressions derived here and the large $K$ results of \cite{cohen13} can be extended to the more general weighted covariance model. This result is general and not specific to the microscopy setting. 

\subsubsection{Low SNR}
\label{low}
\begin{center}
\begin{figure}[t]
\includegraphics[width=80mm]{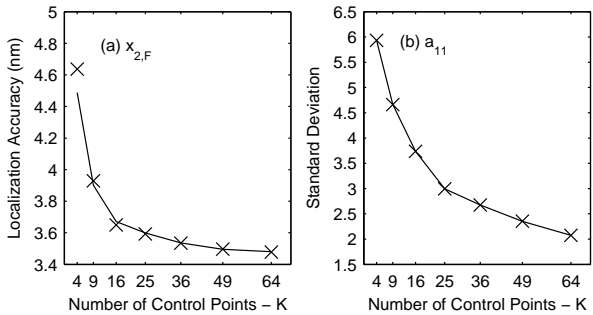}
 \caption{\label{figsSNR}Low SNR example (see Section \ref{low} for more details). The line indicates the square root of the CRLB and the crosses mark the sample standard deviation plotted as a function of the number of CPs $K$ for the following estimated parameters: (a) the first dimension of the unknown single molecule position parameter $x_{2,F}$, (b) $a_{11}$, the leading element of transformation matrix $A$. In this simulation study CPs are in a grid configuration and the transformation consists of a rotation and translation. The vertical axis in (a) has units as nanometers, the vertical axis in (b) is unitless and on the scale $\times 10^{-4}$. Results are based on a sample of $10^6$ simulations.}
\end{figure}
\end{center}

To demonstrate that the CRLB is an appropriate bound for low signal strengths we consider the same simulation study as in Section \ref{rotation} but where the photon count associated with each control point is a uniformly distributed random variable on the interval $[200,700]$ and $300$ photons are collected for the single molecule. Figure \ref{figsSNR} shows the CRLB is still appropriate in this setting.
\subsubsection{Estimating the CRLB}
For the simulations studies presented thus far the theoretical values of the CRLB have been possible to calculate due to artificial knowledge of the true parameter values that form the parameter vector $\theta_{FTC}$. As this is the very thing that needs estimating the theoretical values of the CRLB is obviously unavailable to experimenters and therefore it becomes important to know how well we can estimate the CRLB given the estimated parameter values.

\begin{figure}[t]
\begin{center}$
\begin{array}{cc}
\includegraphics[width=40mm]{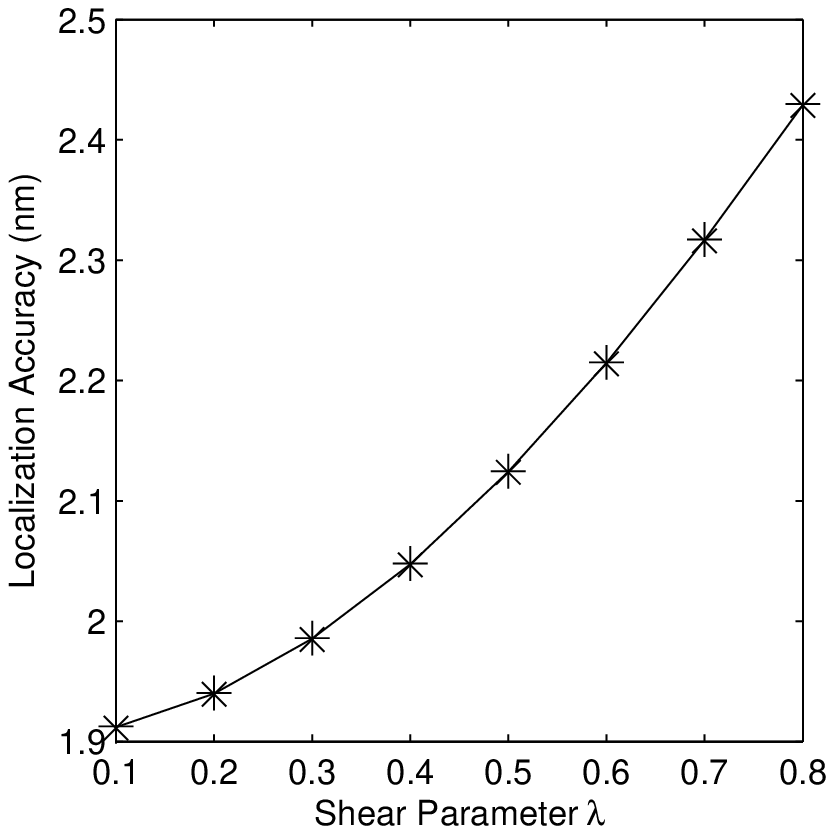}&
\includegraphics[width=40mm]{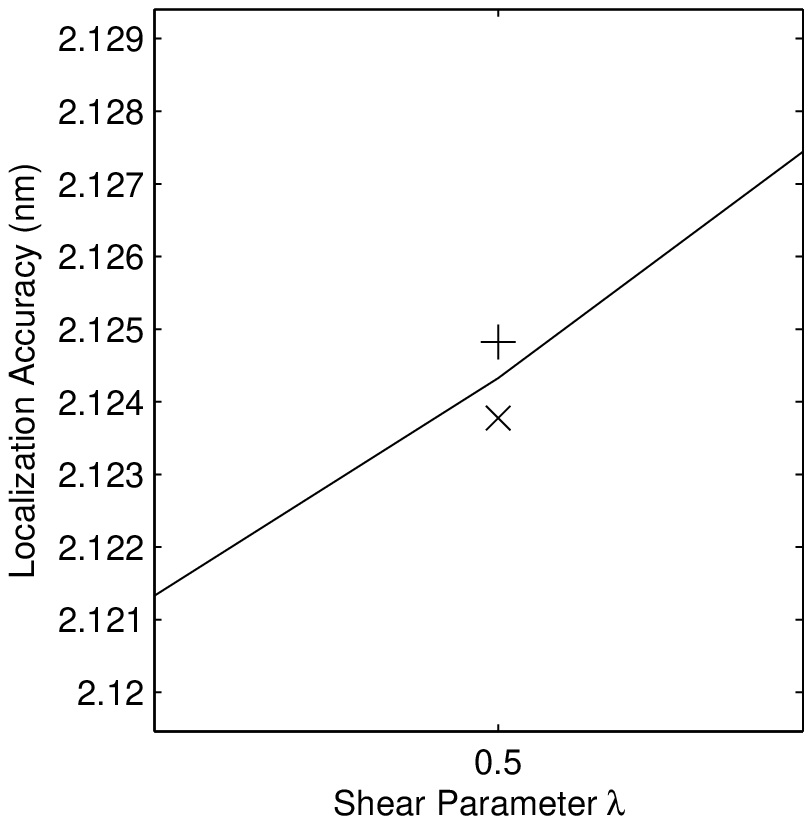}
\end{array}$
\caption{\label{estimated}Left --- the line indicates the square root of the true CRLB for the first dimension of the unknown single molecule position parameter $x_{2,F}$ (the leading diagonal in (\ref{finaleq})) in object space dimensions, plotted as a function of shear parameter $\lambda$. The `$\times$' marks the minimum value of the estimated CRLB and the `+' marks the maximum value of the estimated CRLB taken over $10^6$ simulations. Right --- a magnification of the left plot for a single value of shear parameter.}
\end{center}
\end{figure}

 We consider the same simulation set-up of Section \ref{shear} and now estimate the CRLB from (\ref{finaleq}) using estimated values $\hat{A}$, $\hat{s}$, $y_{1,F}$ and $\{y_{1,1},...,y_{1,K}\}$, instead of true values $A$, $s$, $x_{2,F}$ and $\{x_{1,1},...,x_{1,K}\}$, respectively. In Figure $\ref{estimated}$ we plot the theoretical value of the CRLB for estimating $x_{2,F}$, together with the maximum and minimum value of the estimated CRLB over $10^6$ simulations, clearly demonstrating that estimated parameter values can be used by experimenters to get an excellent estimate of the CRLB.
\subsection{Experimental verification}
Here we describe the experimental set up used to verify the theoretical results of this paper. A bead sample was prepared by adsorbing a dilute solution of 100-nm Tetraspeck microspheres (Thermo Fisher, Waltham, MA, USA) on Poly-L-Lysine (Sigma-Aldrich, St. Louis, MO, USA) coated glass coverslip (Zeiss, Thornwood, NY, USA). A standard inverted microscope (Zeiss Axiovert 200) was configured with a 63$\times$1.46 numerical aperture Zeiss Plan Apochromat objective lens. The beads were excited by a 488nm diode laser (Toptica, Victor, NY, USA) and  a 635nm diode laser (OptoEngine, Midvale, UT, USA). The emission light from the beads was split into two wavelength ranges, 502.5nm-537.5nm and 657.5nm-694.5nm, using a dichroic filter set  (FF560-Di01-25x36; FF01-520/35-25; FF01-676/37-25; Semrock, Rochester, NY, USA), and imaged using two identical charge-coupled device (CCD) cameras (iXon DU897-BV; Andor, South Windsor, CT, USA). 

The imaging experiments were carried out by illuminating the beads with two lasers in 100ms pulse width over 599 repeat acquisitions. To estimate the coordinates of the beads acquired from each camera, we first selected region of interests (ROIs) containing a bead and fitted a Gaussian model using maximum likelihood estimation. All computations were performed using custom written software in MATLAB (MathWorks, Natick, MA, USA). 

\begin{figure}[t]
\begin{center}$
\begin{array}{cc}
\includegraphics[width=40mm]{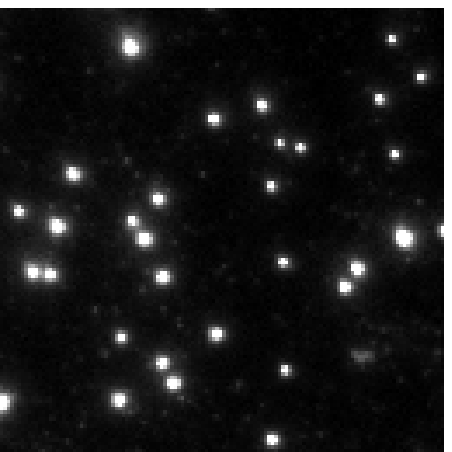}&
\includegraphics[width=40mm]{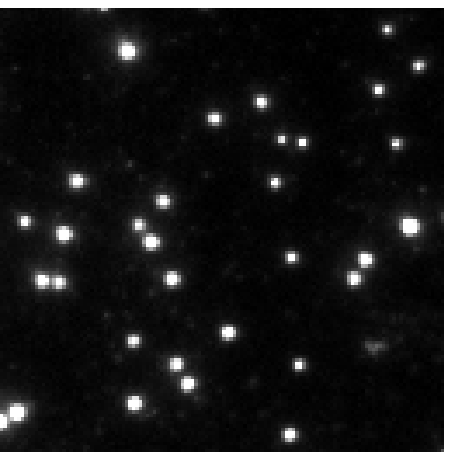}
\end{array}$
\end{center}
\caption{\label{realimage}Example (acquisition 599) of the pair of images of the bead sample to be registered, taken with the two separate cameras as described in Section VI-B.}
\end{figure}

Acquisitions 300-599 (a total of 300) were used in our analysis as they showed the greatest stability in photon counts between acquisitions and hence localization errors are considered approximately iid. An example pair of images from each camera that need to be registered are shown in Figure \ref{realimage}.
\subsubsection{Verification of Gaussian distributed measurement errors}
\begin{figure*}[t!]
\begin{center}
\includegraphics[width=150mm]{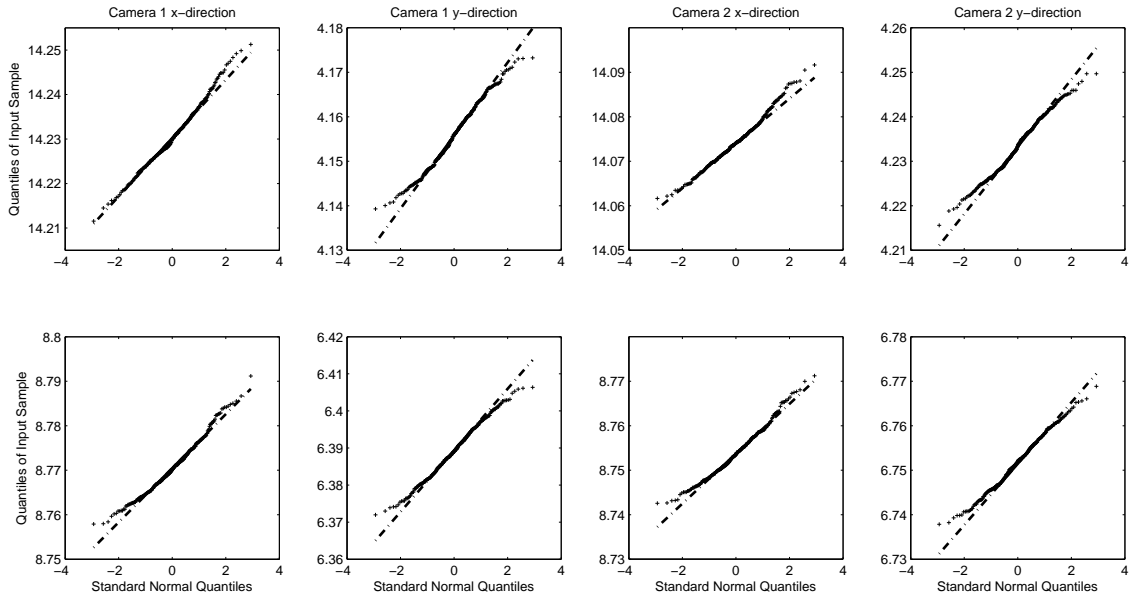}
\end{center}
\caption{\label{figqq} QQ-plots (see Section IV-B1 for details) for the $x$ and $y$ coordinates of the localization estimates in both cameras. Each row of plots corresponds to one of two beads analysed.}
\end{figure*}

Throughout this paper localization errors have been assumed Gaussian in order to form the likelihood function from which the CRLBs are derived. Here, we verify this assumption by analysing the empirical localization estimates for the beads. Figure \ref{figqq} shows quantile-quantile (QQ) plots for the distribution of the localization estimates. The curve is produced by ordering the 300 independent estimates for either the $x$ or $y$ localization coordinate into increasing order of size. The probability of a value less than the $j$th ordered estimate (sample quantile) is approximately $p_j=j/301$. The corresponding theoretical quantile of the standard normal distribution is the value $t_j$ such that $p_j = F(t_j)$, where $F(\cdot)$ is the cumulative density function of the standard normal distribution. The values $t_1,...,t_{300}$ are plotted on the horizontal axis against the ordered estimates on the vertical axes. This is done for $x$ and $y$ coordinates in both cameras for two separate beads (one row of QQ plots for each bead). The straight line indicates the ideal fit for Gaussian samples. It is clear the localization estimates are Gaussian distributed to a close approximation, except for some minor deviations at the distribution tails.
\subsubsection{Registration performance}
\begin{figure}[t]
\begin{center}
\includegraphics[width=80mm]{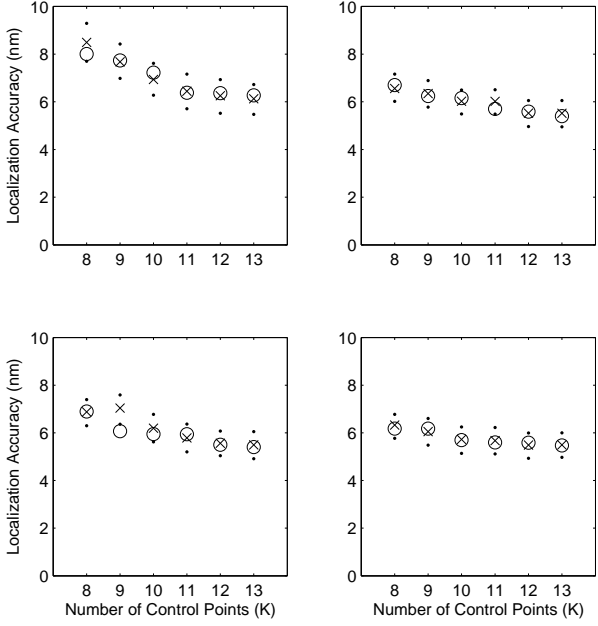}
\end{center}
\caption{\label{exCRLB}Experimental data results. The crosses indicate the square root of the sample variance of the first dimension of the LRE. The dots are the bootstrapped 95\% confidence interval. The circles indicate the square root of the CRLB in the first dimension as calculated from (\ref{eqCRLBLRE}) using the estimated transformation parameters.}
\end{figure}

Fourteen of the fluorescent beads that were present in the field of view for all acquisitions and able to be pair-matched were considered for image registration. One of the beads was isolated as a feature and registration performed using the weighted covariance generalized least squares estimators for $A$ and $s$ \cite{cohen13} with 8,9,10,11,12 and 13 of the remaining beads. Calculating the sample variance of the LRE in the $x$-direction across the 300 acquisitions, we compare it to the CRLB as given in (\ref{eqCRLBLRE}) estimated using the transformation parameter estimates (example values of these are $\hat{A} = \left[\begin{smallmatrix}
0.997 & 0.054 \\ 0.055 & 0.996
\end{smallmatrix}\right]$ and $\hat{s} = [1.000,1.000]^T$ (each to 3 d.p.)). Figure \ref{exCRLB} displays the results. The four plots correspond to four random permutations of the beads we register with. It is clear that in this experiment the CRLB, to a close approximation, is attained.

\section{Concluding remarks}
We have derived the CRLB for image registration performance under the heteroscedastic multivariate errors-in-variables model. Particular focus has been given to the case where the covariance matrices for the errors in localizing the CPs are all scalar multiples of a common positive definite matrix, a suitable model for fluorescence microscopy. Under this model the CRLB for estimating the location of a feature/single molecule has been found and is equal to the lower bound of the covariance matrix of the LRE, the error in localizing a feature/single molecule in the registered image. In the simplified case of that common matrix being the identity and the affine transformation between the pair of images being a scaled version of a unitary matrix, it has been shown that the lower bound for the covariance matrix of the LRE exactly matches the previously published large $K$ expression when transformation parameters are estimated with the weighted covariance generalized least squares estimators. Therefore (\ref{finaleq}) can now be considered to be the lower bound for the accuracy with which we can localize a single molecule in a registered image. Beyond this, it could also be used in future to develop strategies for the placement of the control points so that the estimation errors can be reduced.

Simulations comparing the sample standard deviation of the transformation parameters and an element of the LRE with their respective theoretical lower bounds confirm that using the weighted covariance generalized least squares estimators for the affine parameters appears to be efficient even for low numbers of control points. Experimental data validates the theory presented.


\appendices
\section{}
\label{CRLBTT}
Here we derive the FIM for the parameter vector $\theta_{TC}$ under the most general heteroscedastic model. We note that under the affine transformation assumed then $\mu_{k}\in\mathbb{R}^{2d}$, the mean vector for the measured $k$th CP locations $y_{k} = [y_{1,k}^T,y_{2,k}^T]^T$ is given as
$$
\bmu_k = \left[\begin{array}{c} x_{1,k} \\ Ax_{1,k} + s \end{array} \right].
$$
Therefore we have
$$
\frac{\partial \bmu_k}{\partial \btheta_{TC}^T} = \left[\begin{array}{cc}0 & F_k \\ H_k & G_k \end{array}\right],
$$
where $F_k  = (e_K^{(k)})^T\otimes I_d$, $G_k = (e_K^{(k)})^T\otimes A$, $H_k = \left[I_d\otimes x_{1,k}^T,I_d\right]$.
From (\ref{JTC}) we have 
\begin{align}
J(\theta_{TC}) 
& = \sum_{k=1}^{K} \left[\begin{array}{cc}0 & H^T_k \\ F^T_k & G^T_k \end{array}\right]\left[\begin{array}{cc}\Omega_{1,k}^{-1} & 0 \\ 0 & \Omega_{2,k}^{-1} \end{array}\right] \left[\begin{array}{cc}0 & F_k \\ H_k & G_k \end{array}\right]\nonumber \\
& =\sum_{k=1}^{K} \left[\begin{array}{cc}H_k^T\Omega^{-1}_{2,k}H_k & H^T_k\Omega_{2,k}^{-1}G_k \\ G^T_k\Omega_{2,k}^{-1}H_k & F_k^T\Omega_{1,k}^{-1}F_k + G_k^T \Omega_{2,k}^{-1}G_k \end{array}\right].\nonumber\\
&\equiv \left[\begin{array}{cc}S_{HH} & S_{HG} \\ S_{HG}^T & S_{FF} + S_{GG} \end{array}\right],\nonumber
\end{align}
where $S_{HH} = \sum_{k=1}^{K}H_k^T\Omega_{2,k}^{-1}H_k$, $S_{HG} = \sum_{k=1}^{K}H^T_{k}\Omega_{2,k}^{-1} G_k$, $S_{FF} = \sum_{k=1}^{K}F_k^T\Omega^{-1}_{1,k}F_k$ and $S_{GG} = \sum_{k=1}^K G_k^T\Omega_{2,k}^{-1}G_k$. Dealing with each term individually, we can write 
\begin{multline}
S_{HH}  = \\ \sum_{k=1}^{K}\left[\begin{array}{cccc}
X_{1,k}^T\Omega_{2,k}^{-1} X_{1,k} & \cdots & X_{1,k}^T\Omega_{2,k}^{-1} X_{d,k} & X_{1,k}^T \Omega_{2,k}^{-1} \\ \vdots & & \vdots & \vdots \\
X_{d,k}^T\Omega_{2,k}^{-1} X_{1,k} & \cdots & X_{d,k}^T\Omega_{2,k}^{-1} X_{d,k} & X_{d,k}^T \Omega_{2,k}^{-1} \\ 
\Omega_{2,k}^{-1} X_{1,k}      & \cdots & \Omega_{2,k}^{-1} X_{d,k}      & \Omega_{2,k}^{-1}
\end{array}\right],\label{HH}
\end{multline}
\begin{multline}
S_{HG}  = \\ \left[\begin{array}{cccc}
X_{1,1}^T\Omega_{2,1}^{-1} A^T & X_{1,2}^T\Omega_{2,2}^{-1} A^T & \cdots & X_{1,K}^T \Omega_{2,K}^{-1}A^T \\ \vdots & \vdots & & \vdots \\
X_{d,1}^T\Omega_{2,1}^{-1} A^T & X_{d,2}^T\Omega_{2,2}^{-1} A^T & \cdots & X_{d,K}^T \Omega_{2,K}^{-1}A^T \\ 
\Omega_{2,1}^{-1} A^T      & \Omega_{2,2}^{-1} A^T  & \cdots    & \Omega_{2,K}^{-1}A^T
\end{array}\right],\label{HG} 
\end{multline}
\begin{equation}
S_{FF} + S_{GG} 
 = \left[\begin{array}{cccc}\Lambda^T\Omega_{1}^{-1}\Lambda & 0 & \cdots & 0 \\
0 & \Lambda^T\Omega_{2}^{-1}\Lambda & \cdots & 0 \\
\vdots & \vdots & \ddots & \vdots \\
0 & 0 & \cdots & \Lambda^T\Omega_{K}^{-1}\Lambda
\end{array}\right],\label{FFGG}
\end{equation}
where $\Lambda = [I_d,A^T]^T$ and $X_{i,k} = e_d^{(i)}\otimes x_{1,k}^T$, $i=1,...,d$ and $k=1,...,K$.
\section{}
\label{FIMLRE}
Here we derive the FIM for the parameter vector $\theta_{FTC}$ under the most general heteroscedastic model. The FIM is defined as 
\begin{multline*}
J(\theta_{FTC}) \equiv E\left\{\frac{\partial \mathcal{L}(\btheta_{FTC},x_{2,F}|y_{1},...,y_{K},y_{1,F}) }{\partial  \theta_{FTC}}\right. \\ \left. \times \frac{\partial \mathcal{L}(\btheta_{FTC},x_{2,F}|y_{1},...,y_{K},y_{1,F}) }{\partial \theta_{FTC}^T}\right\}.
\end{multline*}
Given (\ref{loglike}) this can be expressed as
\begin{equation}
\label{combinedFIM}
J(\theta_{FTC}) = \left[\begin{array}{cc}
0 & 0 \\ 
0 & J(\theta_{TC})
\end{array} \right] + J_{F}(\theta_{FTC}),
\end{equation}
where $J(\theta_{TC})$ is as given in (\ref{JTC}) and $$J_F(\theta_{FTC}) = \sum_{k=1}^{K}\frac{\partial \bmu_{1,F}^T}{\partial \btheta_{FTC}}\Omega_{1,F}^{-1}\frac{\partial \bmu_{1,F}}{\partial \btheta_{FTC}^T}.$$ The zeros in the right-hand-side of (\ref{combinedFIM}) are a consequence of the control point localizations being independent of the feature location.
We have the following identities
\begin{align}
\frac{\partial \mu_{1,F}}{\partial a_{ij}} &= -A^{-1}P^{(ij)}A^{-1}(x_{2,F}-s) = -A^{-1}P^{(ij)}x_{1,F}, \nonumber\\
\frac{\partial \mu_{1,F}}{\partial s} &= - A^{-1},\nonumber \\
\frac{\partial \mu_{1,F}}{\partial x_{1,k}} &= {\mathbf 0},\nonumber
\end{align}
where $P^{(ij)}$ is a $d\times d$ matrix of zeros except for a 1 placed in the $(i,j)$th element. It follows from these identities that 
\begin{align}
D_F\equiv \frac{\partial \bmu_{1,F}}{\partial \btheta^T_F} & =  -A^{-1},\nonumber\\
D_T \equiv \frac{\partial \bmu_{1,F}}{\partial \btheta^T_T} & = -A^{-1}\left[x_{1,F}^T\otimes I_d,I_d\right].\nonumber
\end{align}
Define $D_{TT} \equiv D_T^T\Omega_{1,F}^{-1}D_T$ and $D_{FT} \equiv D_F^T\Omega_{1,F}^{-1}D_T$ and $D_{TF} \equiv D_T^T\Omega_{1,F}^{-1}D_F = D_{FT}^T$.
It follows that
\begin{align}
J_F(\theta_{FTC}) & = \frac{\partial \bmu_{1,F}^T}{\partial \btheta_{FTC}}\Omega_{1,F}^{-1}\frac{\partial \bmu_{1,F}}{\partial \btheta^T_{FTC}}\nonumber \\
& = \left[\begin{array}{ccc}
A^{-T}\Omega_{1,F}^{-1}A^{-1} & D_{FT} & 0 \\ D_{FT}^T & D_{TT} & 0 \\ 0 & 0 & 0
\end{array}\right].\nonumber
\end{align}
The zeros in the final row and column can be interpreted as arising because estimating the feature location in $\mathcal{I}_1$ occurs before registration and therefore has no dependence on the CP locations. The expression in (\ref{eqFIMLRE}) follows from (\ref{combinedFIM}).
\section{}
\label{CRLBLRE}
We consider the CRLB block matrix
$$
C(\theta_{FTC})\equiv J^{-1}(\theta_{FTC}) = \left[\begin{array}{ccc}
C_{FF} & C_{FT} & C_{FC} \\
C_{TF} & B_{TT} & B_{TC} \\
C_{CF} & B_{CT} & B_{CC}
\end{array}\right]
$$
where we initially use the notations $B_{TT}$, $B_{TC}$, $B_{CT}$ and $B_{CC}$ to distinguish these from the matrices $C_{TT}$, $C_{TC}$, $C_{CT}$ and $C_{CC}$ considered in (\ref{cthetatc}) and (\ref{eqCRLBTT}).

We note that 
\begin{align}
\left[\begin{array}{cc}B_{TT} & B_{TC} \\
 B_{CT} & B_{CC}\end{array}\right] &= \left(\left[\begin{array}{cc} D_{TT}+ S_{HH} &S_{HG} \\ S_{HG}^T & S_{FF} + S_{GG} \end{array}\right]\right. \nonumber \\ & \hspace{1cm} \left. - \left[\begin{array}{c}D_{FT}^T \\ 0\end{array}\right]A\Omega_{1,F}A^{T}[D_{FT},0]\right)^{-1}\nonumber\\
 & = \left(\left[\begin{array}{cc} D_{TT}+ S_{HH} & S_{HG} \\ S_{HG}^T & S_{FF} + S_{GG} \end{array}\right] \right. \nonumber \\ & \hspace{1cm} \left. - \left[\begin{array}{cc}D_{FT}^T A\Omega_{1,F}A^{T}D_{FT} & 0 \\ 0 & 0\end{array}\right]\right)^{-1}\nonumber.
\end{align}
It is straightforward to show that $D_{FT}^T A\Omega_{1,F}A^TD_{FT} = D_{TT}$ and hence
\begin{align}
\left[\begin{array}{cc}B_{TT} & B_{TC} \\
 B_{CT} & B_{CC}\end{array}\right] & = \left[\begin{array}{cc} S_{HH} & S_{HG} \\ S_{HG}^T & S_{FF} + S_{GG} \end{array}\right]^{-1} \nonumber \\ & = \left[\begin{array}{cc}C_{TT} & C_{TC} \\
  C_{CT} & C_{CC}\end{array}\right]\nonumber
\end{align}
recovering the inverse FIM from Section \ref{CRLBatp} in which only the transformation parameters and CP locations are considered, an expected result stemming from the fact that the feature has no involvement in estimating either the parameters and CP locations. 

We are interested in the term $C_{FF}$ whose diagonals are the CRLB for the localization of the feature/molecule in image $\mathcal{I}_2$. It follows that
\begin{multline*}
C_{FF} = \left(A^{-T}\Omega_{1,F}^{-1}A^{-1} \right.\\  \left. - [D_{FT},0]\left[\begin{array}{cc}
D_{TT}+ S_{HH} & S_{HG} \\ S_{HG}^T & S_{FF} + S_{GG}
\end{array}\right]^{-1}\left[\begin{array}{c}D_{FT}^T \\ 0\end{array}\right]\right)^{-1}\\
 = \left(A^{-T}\Omega_{1,F}^{-1}A^{-1} - D_{FT}M_{11}D_{FT}^T\right)^{-1} \nonumber
\end{multline*}
where
$$
\left[\begin{array}{cc}
M_{11} & M_{12} \\ M_{21} & M_{22}
\end{array}\right] = \left[\begin{array}{cc}
D_{TT}+ S_{HH} & S_{HG} \\ S_{HG}^T & S_{FF} + S_{GG}
\end{array}\right]^{-1}.
$$
We therefore recognise that we can write $C_{FF}$ as in (\ref{eqCRLBLRE}).

\section{}
\label{simplify}
Here we derive $C_{TT}$, the CRLB matrix for estimating the transformation parameters, under Assumption I. Consider each term in (\ref{eqCRLBTT}) with $\Omega_{1,0} = \sigma_{1,0}^2 I_2$, $\Omega_{2,0} = \sigma_{2,0}^2 I_2$ and $A = \varsigma R$, where $R$ is a unitary matrix (rotation/reflection) and $\varsigma \in \mathbb{R}^+$ is a scaling factor. Then from (\ref{HH}), (\ref{HG}) and (\ref{FFGG}) we have
$$
S_{HH} =\sum_{k=1}^K\frac{1}{\sigma_{2,k}^2}\left[\begin{array}{ccc}
\chi_k & 0 & X_{1,k}^T \\ 0 & \chi_k & X_{2,k}^T \\ X_{1,k} & X_{2,k} & I_2
\end{array}\right],
$$
where $\chi_{k} = x_{1,k}x_{1,k}^T,$
\begin{multline*}
S_{HG} = \\ \left[\begin{array}{cccc}
\sigma_{2,1}^{-2}X_{1,1}^T A^T & \sigma_{2,2}^{-2}X_{1,2}^T A^T & \cdots & \sigma_{2,K}^{-2}X_{1,K}^T A^T \\ 
\sigma_{2,1}^{-2}X_{2,1}^T A^T & \sigma_{2,2}^{-2}X_{2,2}^T A^T & \cdots & \sigma_{2,K}^{-2}X_{2,K}^T A^T \\ 
\sigma_{2,1}^{-1} A^T      & \sigma_{2,2}^{-2} A^T  & \cdots    & \sigma_{2,K}^{-2}A^T
\end{array}\right]
\end{multline*}
and
\begin{multline*}
S_{FF} + S_{GG} =  \\ \left[\begin{array}{ccc}
(\sigma_{1,1}^{-2} + \varsigma^2\sigma_{2,1}^{-2}) & \cdots & 0 \\
\vdots & \ddots & \vdots \\
0 & \cdots & (\sigma_{1,K}^{-2} + \varsigma^2\sigma_{2,K}^{-2}) \end{array}\right]\otimes I_2.
\end{multline*}
This gives
\begin{multline} S_{HG}\left(S_{FF} + S_{GG}\right)^{-1}S_{HG}^T \\ = 
\sum_{k=1}^{K}\frac{\varsigma^2}{\sigma_{2,k}^{4}(\sigma_{1,k}^{-2} + \varsigma^2\sigma_{2,k}^{-2})}\left[\begin{array}{ccc}
\chi_k & 0 & X_{1,k}^T \\ 0 & \chi_k & X_{2,k}^T \\ X_{1,k} & X_{2,k} & I_2
\end{array}\right]\nonumber
\end{multline}
and therefore
$$S_{HH} -  S_{HG}\left(S_{FF} + S_{GG}\right)^{-1}S_{HG}^T$$ equals
\begin{multline*}
\sum_{k=1}^{K}\left(\sigma_{2,k}^{-2} -\frac{\varsigma^2}{\sigma_{2,k}^{4}(\sigma_{1,k}^{-2} +\varsigma^2 \sigma_{2,k}^{-2})}\right)\left[\begin{array}{ccc}
\chi_k & 0 & X_{1,k}^T \\ 0 & \chi_k & X_{2,k}^T \\ X_{1,k} & X_{2,k} & I_2
\end{array}\right] \\ = \left(\varsigma^2\sigma_{1,0}^{2} + \sigma_{2,0}^{2}\right)^{-1}\sum_{k=1}^{K}\eta_{k}^{-1}\left[\begin{array}{ccc}
\chi_k & 0 & X_{1,k}^T \\ 0 & \chi_k & X_{2,k}^T \\ X_{1,k} & X_{2,k} & I_2
\end{array}\right]
\end{multline*}
and 
$$C_{TT} = \left(S_{HH} -  S_{HG}\left(S_{FF} + S_{GG}\right)^{-1}S_{HG}^T \right)^{-1}$$ becomes
\begin{multline*} 
C_{TT} = \left(\varsigma^2\sigma_{1,0}^{2} + \sigma_{2,0}^{2}\right)\\\times\left[\begin{array}{ccc}
\Psi^{-1} & 0 & -\Gamma_1^T \\
0 & \Psi^{-1} & -\Gamma_2^T \\
-\Gamma_1 &  -\Gamma_2 & \gamma^{-1} I_2 + \gamma^{-1}\left(\Gamma_1\bar{X}_1^{T}+\Gamma_2\bar{X}_2^{T}\right)
\end{array}\right],
\end{multline*}
where $\gamma \equiv (1/K)\sum_{k=1}^K\eta_k^{-1}$, $\Psi \equiv \Xi - \gamma^{-1}\bar{x}_1\bar{x}_1^T$ where $\Xi\equiv (1/K)\sum_{k=1}^K \eta_{k}^{-1}x_{1,k}x_{1,k}^T$ and $\bar{x}_1\equiv (1/K)\sum_{k=1}^{K}\eta_{k}^{-1}x_{1,k}$, and $\Gamma_i \equiv \gamma^{-1}\bar{X}_i\Psi$ where $\bar{X}_i\equiv \sum_{k=1}^{K}\eta_{k}^{-1} X_{i,k} = e_{2}^{(i)}\otimes \bar{x}_{1}^T$, $i=1,2$.
\section{}
\label{newapp}
Here we consider $C_{FF}$, the CRLB for estimating the location of the feature/single molecule in the registered image, under Assumption I. Appendix \ref{simplify} shows that
$$C_{TT}^{-1} = \left(\varsigma^2\sigma_{1,0}^{2} + \sigma_{2,0}^{2}\right)^{-1}\sum_{k=1}^{K}\eta_{k}^{-1}\left[\begin{array}{ccc}
\chi_k & 0 & X_{1,k}^T \\ 0 & \chi_k & X_{2,k}^T \\ X_{1,k} & X_{2,k} & I_2
\end{array}\right],
$$
under the weighted covariance model and with $\Omega_{1,0} = \sigma_{1,0}^2 I_2$, $\Omega_{2,0} = \sigma_{2,0}^2 I_2$, $\Omega_{1,F} = \sigma_{1,F}^2I_2$ and $A = \varsigma R$ where $R$ is a unitary matrix (rotation/reflection) and $\varsigma \in \mathbb{R}^+$ is a scaling factor, it follows that
\begin{align}
D_{TT} & = \frac{1}{\varsigma^2\sigma_{1,F}^2}\left[\begin{array}{cc}
x_{1,F}x_{1,F}^T & x_{1,F} \\ 
x_{1,F}^T & 1
\end{array} \right]\otimes I_2,\nonumber\\
D_{FT} & = \frac{1}{\varsigma^2\sigma_{1,F}^2}\left[x_{1F}^T,1\right]\otimes I_2.\nonumber
\end{align}
Therefore the result follows from (\ref{eqCRLBLRE}).
\section{}
\label{CRLBsupersimple}
Here we derive $C_{FF}$ under Assumptions I and II. With $(1/K)\sum_{k=1}^{K}\eta_k^{-1}\chi_k = \nu^2 I_2$ and $(1/K)\sum_{k=1}^{K}\eta_k^{-1}X_{j,k} = 0$ (\ref{CFFeq}) becomes
\begin{multline*}
C_{FF} = \left(\alpha^{-1}I_2 - \right. \\ \left.  \alpha^{-2}\left[x_{1F}^T,1\right]\otimes I_2\left(\alpha^{-1}\left[\begin{array}{cc}
x_{1,F}x_{1,F}^T & x_{1,F} \\ 
x_{1,F}^T & 1
\end{array} \right]\otimes I_2  \right.\right.  \\ \left.\left.   + \beta^{-1}\left[\begin{array}{cc}
K\nu^2 I_2 & 0 \\ 0 & K\gamma
\end{array}\right]\otimes I_2\right) ^{-1}\left[x_{1F}^T,1\right]^T\otimes I_2\right)^{-1}
\\ = \left(\alpha^{-1}I_2 - \alpha^{-2}\left[x_{1F}^T,1\right]\otimes I_2 \left(L^{-1}\otimes I_2 \right) \left[x_{1F}^T,1\right]^T\otimes I_2 \right)^{-1} \\
 = \left(\alpha^{-1}I_2 - \alpha^{-2}\left(\left[x_{1F}^T,1\right]L^{-1} \left[x_{1F}^T,1\right]^T\right)\otimes I_2\right)^{-1}, \nonumber
\end{multline*}
where 
$$
L = \left[\begin{array}{cc}
\alpha^{-1}x_{1,F}x_{1,F}^T+\beta^{-1}K\nu^2 I_2 & \alpha^{-1}x_{1,F} \\ 
\alpha^{-1}x_{1,F}^T & \alpha^{-1} + K\gamma\beta^{-1}
\end{array} \right],
$$
with $\alpha = \varsigma^2\sigma_{1,F}^2$ and $\beta = \varsigma^2\sigma_{1,0}^{2} + \sigma_{2,0}^{2}$.

Let $A = \alpha^{-1}x_{1,F}x_{1,F}^T+\beta^{-1}K\nu^2 I_2$, $B = \alpha^{-1}x_{1,F}$, $C = \alpha^{-1}x_{1,F}^T$, $D = \alpha^{-1} + K\gamma\beta^{-1}$, then $A  - BD^{-1}C = \alpha^{-1}\Theta+\beta^{-1}K\nu^2 I_2 - \alpha^{-2}(\alpha^{-1}+K\gamma\beta^{-1})^{-1}\Theta = (\alpha + (K\gamma)^{-1}\beta)^{-1}\Theta + \beta^{-1}K\nu^2 I_2$, where $\Theta = x_{1,F}x_{1,F}^T$. If $a = (\alpha + (K\gamma)^{-1}\beta)^{-1}$ and $b = \beta^{-1}K\nu^2$ then $(A  - BD^{-1}C)^{-1} = (a\Theta + bI_2)^{-1} = b^{-1}(b+ar^2)^{-1}\left( a\Theta^\ast + bI_2 \right)$, where $\Theta^\ast = R \Theta R^T$, with $R$ the $\pi/2$ rotation matrix. This gives
$(A - BD^{-1}C)^{-1} = \beta K^{-1}\nu^{-2}(\alpha\beta^{-1}+K\nu^2 + r^2 + K\nu^2)^{-1}\Theta^\ast + cI_2$, where $c = (\beta^{-1}K\nu^2 + (\alpha+(K\gamma)^{-1}\beta)^{-1}r^2)^{-1}$. In a further condensing of notation we define $\bar{\beta} \equiv (1/K)\gamma^{-1}\beta$. With $\Theta^\ast x_{1,F} = 0$ it can be shown that
\begin{align}
C_{FF} 
& = (\alpha^{-1} - \alpha^{-2}\left(cr^2 - 2cr^2(\alpha^{-1}+\bar\beta^{-1})^{-1}\alpha^{-1} + \nonumber \right. \\ & \hspace{0.6cm} \left. (\alpha^{-1}+\bar\beta^{-1})^{-1} + cr^2\alpha^{-2}(\alpha^{-1}+\bar\beta^{-1})^{-2}\right))^{-1} I_2\nonumber \\
& = \alpha(1 - \alpha^{-1}(cr^2(\alpha^{-1}(\alpha^{-1}+\bar\beta^{-1})^{-1}-1)^2+ \nonumber \\ &\hspace{1cm} (\alpha^{-1}+\bar\beta^{-1})^{-1}))^{-1} I_2\nonumber \\
& = \alpha\left(1 - \alpha^{-1}\left(\frac{cr^2\alpha^2}{(\alpha+\bar{\beta})^2} + (\alpha^{-1}+\bar\beta^{-1})^{-1}\right)\right)^{-1} I_2\nonumber \\
& = \alpha\left(1 - \frac{cr^2\alpha}{(\alpha+\bar{\beta})^2} - \frac{\bar{\beta}}{\alpha+\bar{\beta}}\right)^{-1} I_2\nonumber.
\end{align}
With
$$
\frac{cr^2\alpha}{(\alpha+\bar{\beta})^2} = \left(\frac{\nu^2}{\gamma r^2}(\alpha^{-1}+\bar{\beta}^{-1}) + \alpha^{-1}\right)^{-1}(\alpha+\bar{\beta})^{-1}
$$
it follows that
\begin{align}
C_{FF} &= \alpha\left(1 - \frac{\left(\frac{\nu^2}{\gamma r^2}(\alpha^{-1}+\bar{\beta}^{-1}) + \alpha^{-1}\right)^{-1}}{(\alpha+\bar{\beta})} - \frac{\bar{\beta}}{\alpha+\bar{\beta}}\right)^{-1} I_2 \nonumber \\ 
&=
\alpha\left( \frac{\alpha-\left(\frac{\nu^2}{\gamma r^2}(\alpha^{-1}+\bar{\beta}^{-1}) + \alpha^{-1}\right)^{-1}}{\alpha+\bar{\beta}}\right)^{-1} I_2\nonumber \\ & = \alpha(\alpha+\bar{\beta})\left( \alpha-\frac{1}{\left(\frac{\nu^2}{\gamma r^2}(\alpha^{-1}+\bar{\beta}^{-1}) + \alpha^{-1}\right)}\right)^{-1} I_2\nonumber \\
& = \alpha(\alpha+\bar{\beta})\left( \frac{\alpha\frac{\nu^2}{\gamma r^2}(\alpha^{-1}+\bar{\beta}^{-1})}{\left(\frac{\nu^2}{\gamma r^2}(\alpha^{-1}+\bar{\beta}^{-1}) + \alpha^{-1}\right)}\right)^{-1} I_2 \nonumber \\ & = (\alpha+\bar{\beta})\left(1+ \frac{1}{\alpha\left(\frac{\nu^2}{\gamma r^2}(\alpha^{-1}+\bar{\beta}^{-1})\right)}\right) I_2\nonumber\\
& = (\alpha+\bar{\beta})\left(1+ \frac{\frac{\gamma r^2}{ \nu^2}\bar{\beta}}{\alpha + \bar{\beta}}\right) I_2\nonumber \\ 
& = \alpha + \bar{\beta}\left(1+\frac{\gamma r^2}{ \nu^2}\right)I_2 \nonumber \\ 
& = \alpha + \frac{1}{K\gamma}\beta\left(1+\frac{\gamma r^2}{ \nu^2}\right)I_2.\nonumber
\end{align}
\bibliographystyle{unsrt}
\bibliography{references_short}
\end{document}